\documentclass[%
print,
 amsmath,amssymb,
 aps,
]{revtex4}

\usepackage[sc]{mathpazo}
\usepackage{graphicx}
\usepackage{dcolumn}
\usepackage{bm}
\usepackage{dsfont}
\usepackage[landscape,papersize={297.1mm,210mm},left=1.9cm,right=1.6cm,top=2.7cm,bottom=2.8cm]{geometry}
\usepackage[                   
            pdfstartview=FitH,
            colorlinks, 
            pdfborder=001,   
            linkcolor=blue,
            anchorcolor=blue,
            citecolor=blue,
            urlcolor=blue
            ]{hyperref}

\begin{document}
\title{Quantifying the quantumness of ensembles via unitary similarity invariant norms}

\author{Xianfei Qi}
\author{Ting Gao}
\email{gaoting@hebtu.edu.cn}
\affiliation {College of Mathematics and Information Science, Hebei
Normal University, Shijiazhuang 050024, China}
\author{Fengli Yan}
\email{flyan@hebtu.edu.cn}
\affiliation {College of Physics Science and Information Engineering, Hebei
Normal University, Shijiazhuang 050024, China}

\begin{abstract}
The quantification of the quantumness of a quantum ensemble has theoretical and practical significance in quantum information theory. We propose herein a class of measures of the quantumness of quantum ensembles using the unitary similarity invariant norms of the commutators of the constituent density operators of an ensemble. Rigorous proof shows that they share desirable properties for a measure of quantumness, such as positivity, unitary invariance, concavity under probabilistic union, convexity under state decomposition, decreasing under coarse graining, and increasing under fine graining. Several specific examples illustrate the applications of these measures of quantumness in studying quantum information.
\end{abstract}


\maketitle

\section{Introduction}
In classical physics, the state space of a system is the phase space. The pure state of an individual system is described by the phase point, while the mixed state is described by a normalized distribution function. In quantum physics, the state space of a quantum system is a Hilbert space. The pure state of a quantum system is described by the state vector, which is a complex unit vector. The mixed state is described by the density operator, which is a trace-one semipositive operator \cite{QCI2010,QI2011}. A mixed quantum state $\rho$ can be prepared (or decomposed) as the mixture of a family of quantum states $\rho_{i}$ with probability $p_{i}$. In the language of mathematics, it means that a quantum ensemble $\{(p_{i},\rho_{i})|i\in I\}$, which consists of a number of states $\rho_{i}$ ($i$ is an index) with respective probabilities $p_{i}$, is represented by a density operator.

A quantum ensemble naturally induces a unique density operator. However, a density operator generally induces many quantum ensembles because of the fact that a mixed quantum state can be prepared as the mixture of quantum states in many different ways (i.e., many ensemble decompositions exist for a density operator of a mixed state). Different ensembles may have different properties. For instance, an unknown state from ensemble $\mathcal{E}_{ort}$ consisting of orthogonal pure states could be perfectly cloned and determined
without being disturbed, while a state from ensemble $\mathcal{E}_{non}$ consisting of non-orthogonal states cannot be perfectly cloned and exactly determined \cite{PRL74.1259}. Intuitively, the quantum characteristics exhibited by ensemble $\mathcal{E}_{non}$ are more than those of $\mathcal{E}_{ort}$, which leads us to focus on the quantumness of ensembles. A question then naturally arises: how does one quantify the quantumness of a quantum ensemble?

Some studies have been conducted on the issue of quantifying the quantumness of quantum ensembles. In \cite{QCCM2002,QIC2003,0302108}, as earlier researchers in this field, Fuchs et al. proposed the measures of the quantumness of an ensemble in terms of the difficulty of transmitting the states through a classical communication channel. They also discussed the potential applications of measures in quantum eavesdropping. A new measure \cite{IJQI2006} of quantumness of an ensemble was presented using the difference between the information content in one copy and the information content in two or more copies. The definition of trace distance and fidelity measure between quantum ensembles was given in \cite{PRA79.032336}, where the operational interpretations for measures were also presented. A measure of the quantumness of an ensemble was defined in \cite{PLA2011} based on the duality relation with a quantity characterizing how
classical a quantum ensemble is. Accordingly, a measure for characterizing the quantumness of an ensemble was introduced from the perspective of studying the local indistinguishability problem of quantum states \cite{SR2014}. Luo et al. paid persistent efforts on this issue and proposed some measures of quantumness via the relative entropy between quantum ensembles \cite{PMH2009} through the disturbances induced by the von Neumann measurement \cite{QIP2010} and commutators \cite{TMP2011,PRA96.022132}.

The present study investigates the problem of quantifying the quantumness of quantum ensembles from a different perspective. Norms are useful mathematical tools in quantum information theory. For example, one can efficiently describe and quantify quantum discord \cite{PRL105.190502} and measure-induced nonlocality (MIN) \cite{PRL106.120401,NJP2015} using different norms. Moreover, notice that the constituent density operators of an ensemble do not generally commute with each other. Combining the two abovementioned points, we propose herein a class of measures of the quantumness of quantum ensembles by virtue of some special norms of the commutators of the constituent density operators of an ensemble. A detailed introduction of their applications in quantum information is then presented. For example, we find a relation between the measure of quantumness and the coherence measure for the qubit pure state. We also provide a relation between the quantumness measure and the entanglement concurrence for two-qubit pure states.

The remainder of this paper is organized as follows: Sec.~\uppercase\expandafter{\romannumeral 2} introduces the definition of the unitary similarity invariant norms (USINs) and lists several famous USINs, such as the Schatten $p$-norms and Ky Fan $p$-norms. We also define a class of measures of quantumness based on USINs and prove that they are valid measures of quantumness. Sec.~\uppercase\expandafter{\romannumeral 3} presents some applications in the quantum information of these measures. Sec.~\uppercase\expandafter{\romannumeral 4} provides the conclusion.

\section{Quantumness measures based on the unitary similarity invariant norms}
Before we state the main results, we will first introduce some necessary knowledge about norms that will be used in the subsequent sections. Throughout the paper, norms are defined on $M_{n}$, where $M_{n}$ denotes the set of all $n$-by-$n$ matrices over complex field $\mathbb{C}$. A norm $\| \cdot\|$ on $M_{n}$ is said to be unitarily invariant if $\| A\|=\| UAV\|$ for all $A\in M_{n}$ and all unitary $U, V\in M_{n}$ \cite{Zhan.2013}. A norm $\| \cdot\|$ on $M_{n}$ is said to be a unitary similarity invariant if $\| A\|=\| UAU^{\dag}\|$ for all $A\in M_{n}$ and all unitary $U\in M_{n}$.

We list some important norms that are unitarily invariant and, thus, are unitary similarity invariant \cite{Zhan.2013}.

\emph{Schatten p-norms}~~~~ Schatten $p$-norms are defined by
$$\| A\|_{p}=\left[\sum\limits_{j=1}^{n}s_{j}^{p}(A)\right]^{1/p},$$
where $s_{j}(A)$ are singular values of A in a nonincreasing order. Here, $\| A\|_{1}$ is the trace norm, which can be expressed as $\| A\|_{1}=\text{tr}\sqrt{AA^{\dag}}$. Meanwhile, $\| A\|_{2}=\| A\|_{F}$ is the Frobenius (or Hilbert--Schmidt) norm that can be expressed as:
$\| A\|_{F}=(\text{tr}A^{\dag}A)^{1/2}=\big(\sum\limits_{i,j}|a_{ij}|^{2}\big)^{1/2}$.

\emph{Ky Fan $p$-norms}~~~~ Ky Fan $p$-norms are defined by
$$\| A\|_{(k)}=\sum\limits_{j=1}^{k}s_{j}(A),$$
where $s_{j}(A)$ are singular values of A in a nonincreasing order. Note that $\| A\|_{(1)}=\| A\|_{\infty}$ and $\| A\|_{(n)}=\| A\|_{1}$, where $\| A\|_{\infty}$ is the spectral norm.

We propose
\begin{equation}
\begin{aligned}
M(\mathcal{E})=\sum\limits_{ij}\sqrt{p_{i}p_{j}}\|[\rho_{i},\rho_{j}]\|,
\end{aligned}
\end{equation}
as a class of measures of the quantumness of the ensemble $\mathcal{E}=\{(p_{i},\rho_{i})|i\in I\}$, where $\|\cdot\|$ denotes a unitary similarity invariant norm.

The constituent density operators of an ensemble do not generally commute with each other, and many quantum features are derived from the noncommutativity between the constituent density operators of an ensemble. Hence, naturally, the commutator of the quantum states should have a place in the definition of the measures of the quantumness of ensembles. Note that the square root of a probability distribution plays a more vital role than the probability distribution itself. Moreover, the unitary similarity invariant norms possess a desirable property, which will be reflected in the following proof of property 2. We establish these measures of quantumness based on the abovementioned ideas.

Next, we prove that these quantumness measures $M(\mathcal{E})$ share the following desirable properties:

(1) (Positivity). $M(\mathcal{E})\geqslant 0$ for arbitrary quantum ensemble $\mathcal{E}$, and equality holds if and only if $\mathcal{E}$ is a classical ensemble. Ensemble $\{(p_{i},\rho_{i})|i\in I\}$ is called classical if all the states in the ensemble commute with each other. Its quantumness vanishes if the ensemble degenerates to a single state.

It follows immediately from the nonnegativity and positivity of the norm.

(2) (Unitary invariance). $M(\cdot)$ is invariant under unitary transformations. That is, $M(\mathcal{E})=M(U\mathcal{E} U^{\dag})$ for any unitary operator $U$. Here, $U\mathcal{E} U^{\dag}=\{(p_{i},U\rho_{i}U^{\dag})|i\in I\}$.

Clearly,
\begin{equation}
\begin{aligned}
M(U\mathcal{E} U^{\dag})&=\sum\limits_{ij}\sqrt{p_{i}p_{j}}\| U\rho_{i}U^{\dag}U\rho_{j}U^{\dag}-U\rho_{j}U^{\dag}U\rho_{i}U^{\dag}\|\\
&=\sum\limits_{ij}\sqrt{p_{i}p_{j}}\| U[\rho_{i},\rho_{j}]U^{\dag}\|\\
&=\sum\limits_{ij}\sqrt{p_{i}p_{j}}\| [\rho_{i},\rho_{j}]\|=M(\mathcal{E}).
\end{aligned}
\end{equation}

(3) (Concavity under probabilistic union). $M(\cdot)$ does not decrease under the mixing of ensembles in the sense that

$$M\left(\bigcup\limits_{\mu}\lambda_{\mu}\mathcal{E}_{\mu}\right)\geqslant \sum\limits_{\mu}\lambda_{\mu}M(\mathcal{E}_{\mu}).$$
Here, $\lambda_{\mu}\geqslant 0$, $\sum_{\mu\in K}\lambda_{\mu}=1$, and for each $\mu\in K$, $\mathcal{E}_{\mu}=\{(p_{\mu i},\rho_{\mu i})|i\in I_{\mu}\}$ is a quantum ensemble, and $\bigcup_{\mu}\lambda_{\mu}\mathcal{E}_{\mu}=\{(\lambda_{\mu}p_{\mu i},\rho_{\mu i})|\mu\in K, i\in I_{\mu}\}$ is a probabilistic union of $\mathcal{E}_{\mu}$.

\emph{Proof:} The probabilistic union of ensembles $\{\mathcal{E}_{\mu}\}$ corresponds to a new ensemble $\mathcal{E}$
$$\mathcal{E}=\bigcup\limits_{\mu}\lambda_{\mu}\mathcal{E}_{\mu}=\{(\lambda_{\mu}p_{\mu i},\rho_{\mu i})|\mu\in K, i\in I_{\mu}\},$$
we then have
\begin{equation}
\begin{aligned}
M(\mathcal{E})&=\sum\limits_{\mu\upsilon ij}\sqrt{\lambda_{\mu}\lambda_{\upsilon}p_{\mu i}p_{\upsilon j}}\| [\rho_{\mu i},\rho_{\upsilon j}]\|\\
&=\sum\limits_{\mu\upsilon ij(\mu=\upsilon)}\sqrt{\lambda_{\mu}\lambda_{\upsilon}p_{\mu i}p_{\upsilon j}}\| [\rho_{\mu i},\rho_{\upsilon j}]\|+\sum\limits_{\mu\upsilon ij(\mu\neq\upsilon)}\sqrt{\lambda_{\mu}\lambda_{\upsilon}p_{\mu i}p_{\upsilon j}}\| [\rho_{\mu i},\rho_{\upsilon j}]\|\\
&\geqslant \sum\limits_{\mu}\lambda_{\mu}\sum\limits_{ij}\sqrt{p_{\mu i}p_{\mu j}}\| [\rho_{\mu i},\rho_{\mu j}]\|=\sum\limits_{\mu}\lambda_{\mu}M(\mathcal{E}_{\mu}).
\end{aligned}
\end{equation}

(4) (Convexity under state decomposition). $M(\cdot)$ does not increase under state decomposition with respect to each constituent state. More specifically, if a constituent state $\rho_{c}$ in quantum ensemble $\mathcal{E}=\{(p_{i},\rho_{i})|i\in I\}$ has the decomposition $\rho_{c}=\sum_{\mu\in J}\lambda_{\mu}\rho_{c\mu}$ with $\sum_{\mu}\lambda_{\mu}=1$ and $\lambda_{\mu}\geqslant 0$, this decomposition induces new ensembles,
$$\mathcal{E}_{\mu}=\{(p_{c},\rho_{c\mu})\}\cup \{(p_{i},\rho_{i})|i\neq c, i\in I\}, \mu\in J,$$
and there is
$$M\left(\sum\limits_{\mu}\lambda_{\mu}\mathcal{E}_{\mu}\right)\leqslant\sum\limits_{\mu}\lambda_{\mu}M(\mathcal{E}_{\mu}).$$

Symbolically,

$$\sum\limits_{\mu}\lambda_{\mu}\mathcal{E}_{\mu}=\left\{\left((p_{c},\sum\limits_{\mu}\lambda_{\mu}\rho_{c\mu})\right)\right\}\cup\{(p_{i},\rho_{i})|i\neq c,i\in I\}=\mathcal{E}.$$

\emph{Proof:} Without loss of generality, we can choose $\rho_{1}$ as $\rho_{c}$. By triangle inequality of the norm, we then obtain
\begin{equation}
\begin{aligned}
M(\mathcal{E})&=\sum\limits_{ij}\sqrt{p_{i}p_{j}}\|[\rho_{i},\rho_{j}]\|\\
&=2\sum\limits_{i>j}\sqrt{p_{i}p_{j}}\| [\rho_{i},\rho_{j}]\|\\
&=2\sum\limits_{i>j(j=1)}\sqrt{p_{i}p_{j}}\|[\rho_{i},\rho_{j}]\|+2\sum\limits_{i>j(j\neq 1)}\sqrt{p_{i}p_{j}}\|[\rho_{i},\rho_{j}]\|\\
&=2\sum\limits_{i>1}\sqrt{p_{i}p_{1}}\| [\rho_{i},\sum\limits_{\mu}\lambda_{\mu}\rho_{1\mu}]\|+2\sum\limits_{\mu}\lambda_{\mu}\sum\limits_{i>j(j\neq 1)}\sqrt{p_{i}p_{j}}\|[\rho_{i},\rho_{j}]\|\\
&\leqslant 2\sum\limits_{i>1}\sqrt{p_{i}p_{1}} \sum\limits_{\mu}\lambda_{\mu}\| [\rho_{i},\rho_{1\mu}]\|+2\sum\limits_{\mu}\lambda_{\mu}\sum\limits_{i>j(j\neq 1)}\sqrt{p_{i}p_{j}}\|[\rho_{i},\rho_{j}]\|\\
&=\sum\limits_{\mu}\lambda_{\mu}\left(2\sum\limits_{i>1}\sqrt{p_{i}p_{1}}\|[\rho_{i},\rho_{1\mu}]\|+2\sum\limits_{i>j(j\neq 1)}\sqrt{p_{i}p_{j}}\| [\rho_{i},\rho_{j}]\|\right)\\
&=\sum\limits_{\mu}\lambda_{\mu}M(\mathcal{E}_{\mu}),
\end{aligned}
\end{equation}
as required.

(5a) (Increasing under fine graining). $M(\cdot)$ increases under fine graining in the sense that $M(\mathcal{E})\leqslant M(\mathcal{E}_{F})$. For each constituent state $\rho_{i}$ in ensemble $\mathcal{E}=\{(p_{i},\rho_{i})|i\in I\}$, it can be decomposed into a new quantum ensemble $\mathcal{E}_{i}=\{(\lambda_{i\mu},\rho_{i\mu})|\mu\in J_{i}\}$, that is, $\rho_{i}=\sum_{\mu\in J_{i}}\lambda_{i\mu}\rho_{i\mu}$, and $\mathcal{E}_{F}=\{(p_{i}\lambda_{i\mu},\rho_{i\mu})|i\in I,\mu\in J_{i}\}$ is called the fine-grained ensemble of $\mathcal{E}$.

\emph{Proof:} Proving the following is not difficult:
\begin{equation}
\begin{aligned}
M(\mathcal{E})&=\sum\limits_{ij}\sqrt{p_{i}p_{j}}\| [\rho_{i},\rho_{j}]\|\\
&=\sum\limits_{ij}\sqrt{p_{i}p_{j}}\| [\sum\limits_{\mu}\lambda_{i\mu}\rho_{i\mu},\sum\limits_{\upsilon}\lambda_{j\upsilon}\rho_{j\upsilon}]\|\\
&\leqslant \sum\limits_{ij}\sqrt{p_{i}p_{j}}\sum\limits_{\mu\upsilon}\lambda_{i\mu}\lambda_{j\upsilon}\| [\rho_{i\mu},\rho_{j\upsilon}]\|\\
&=\sum\limits_{ij\mu\upsilon}\sqrt{p_{i}p_{j}}\lambda_{i\mu}\lambda_{j\upsilon}\| [\rho_{i\mu},\rho_{j\upsilon}]\|\\
&\leqslant \sum\limits_{ij\mu\upsilon}\sqrt{p_{i}\lambda_{i\mu}}\sqrt{p_{j}\lambda_{j\upsilon}}\| [\rho_{i\mu},\rho_{j\upsilon}]\|\\
&=M(\mathcal{E}_{F}).
\end{aligned}
\end{equation}

(5b) (Decreasing under coarse graining). $M(\cdot)$ decreases under coarse graining in the sense that $M(\mathcal{E})\geqslant M(\mathcal{E}_{C})$. Here $\mathcal{E}=\{(p_{i},\rho_{i})|i\in I\}$, and $\{c_{s}|s\in S\}$ is a partition of the index $I$ (i.e., $\{c_{s}|s\in S\}$ is a collection of pairwise disjoint sets and the union $\bigcup_{s}c_{s}=I$). We define $p_{c_{s}}:=\sum_{i\in c_{s}}p_{i}$, $\rho_{c_{s}}:=\frac{1}{p_{c_{s}}}\sum_{i\in c_{s}}p_{i}\rho_{i}$, and $\mathcal{E}_{C}=\{(p_{c_{s}},\rho_{c_{s}})|s\in S\}$ is called the coarse-grained ensemble of $\mathcal{E}$.

\emph{Proof:} ~~Note that ensemble $\mathcal{E}$ is the fine-grained ensemble of ensemble $\mathcal{E}_{C}$, which directly follows from (5a).

\section{Applications}
This section reveals the potential link between the quantumness of the ensembles and some branches of quantum information by considering the following examples:

\emph{Example 1}~~ Consider the quantum ensemble $\mathcal{E}=\{(p_{i},\rho_{i})\}$ ($i=1,2$) that consists of two qubit states. An arbitrary density matrix for a mixed qubit state may be written as $\rho_{i}=\frac{\mathds{1}+\vec{r_{i}}\cdot\vec{\sigma}}{2}$, where $\vec{\sigma}=(\sigma_{x},\sigma_{y},\sigma_{z})$ with $\sigma_{x}, \sigma_{y}, \sigma_{z}$ being Pauli matrices; and $\vec{r_{i}}$ is the Bloch vector for state $\rho_{i}$, which is a real three-dimensional vector such that $|\vec{r_{i}}|\leqslant 1$. We obtain $M(\mathcal{E})=2\sqrt{p_{1}p_{2}}\|[\rho_{1},\rho_{2}]\|_{2}= \sqrt{2p_{1}p_{2}}|\vec{r_{1}}||\vec{r_{2}}|\text{sin}\alpha$ after the computation. Here, $\alpha$ is the angle between vectors $\vec{r_{1}}$ and $\vec{r_{2}}$.

\emph{Example 2}~~ We can investigate the quantumness of ensemble $\mathcal{E}=\{(p_{1},\rho),(p_{2},\Phi(\rho))\}$, where $\rho$ is an input state, and $\Phi(\rho)$ is an output state, to study the decoherent capability of channel $\Phi$. We choose $\Phi$ as the phase damping channel. The Kraus operators for the phase damping channel are expressed as $E_{0}=|0\rangle\langle 0|+\sqrt{1-\lambda}|1\rangle\langle 1|$, $E_{1}=\sqrt{\lambda}|1\rangle\langle 1|$, while $\Phi(\rho)=\sum_{i=0}^{1}E_{i}\rho E_{i}^{\dag}$. If $\rho$ is any pure qubit state with the Bloch vector $\vec{r}=(\text{sin}\theta\text{cos}\phi,\text{sin}\theta\text{sin}\phi,\text{cos}\theta)$, the phase damping channel performs the transformation
$$(\text{sin}\theta\text{cos}\phi,\text{sin}\theta\text{sin}\phi,\text{cos}\theta)\rightarrow (\sqrt{1-\lambda}\text{sin}\theta\text{cos}\phi,\sqrt{1-\lambda}\text{sin}\theta\text{sin}\phi,\text{cos}\theta).$$
According to Example 1, we have $M(\mathcal{E})=2\sqrt{p_{1}p_{2}}\|[\rho,\Phi(\rho)]\|_{2}=\sqrt{2p_{1}p_{2}}(1-\sqrt{1-\lambda})\text{sin}\theta\text{cos}\theta$. $d M(\mathcal{E})/d\lambda \geqslant 0$, which shows that the amount $M(\mathcal{E})$ is an increasing function of $\lambda$ and indicates that this measure can be viewed as a measure of the decoherent capability of the phase damping channel $\Phi$.

\emph{Example 3}~~For an ensemble consisting of two pure $d$ dimensional states $\mathcal{E}=\{(p_{1},|\psi\rangle),(p_{2},|\phi\rangle)\}$ with $\langle\psi|\phi\rangle=ce^{i\theta}(c\in [0,1],\theta\in[0,2\pi))$. Similar to the computation in \cite{SR2014}, we obtain $\|[|\psi\rangle\langle \psi|,|\phi\rangle\langle \phi|]\|_{1}=2c\sqrt{1-c^{2}}$. $M(\mathcal{E})=2\sqrt{p_{1}p_{2}}\|[|\psi\rangle\langle \psi|,|\phi\rangle\langle \phi|]\|_{1}=4c\sqrt{p_{1}p_{2}}\sqrt{1-c^{2}}$. $M_{\mathcal{E}}=0$ if $c=|\langle\psi|\phi\rangle|=0$ or $1$. When $p_{1}=p_{2}=1/2$, $M_{\mathcal{E}}^{max}=1$ if $c=1/\sqrt{2}$ (i.e., when the two states have an angle $45^{\circ}$ between them) \cite{QCCM2002}.

\emph{Example 4}~~Consider the quantum ensemble $\mathcal{E}=\{(1/2,|\psi\rangle),(1/2,|+\rangle)\}$, where $|\psi\rangle$ is a single qubit pure state, $|+\rangle=(|0\rangle+|1\rangle)/\sqrt{2}$. Recall that a single qubit pure state can be written as $|\psi\rangle=\alpha|0\rangle+\beta|1\rangle$ ($\alpha, \beta$ are real numbers) by performing a unitary incoherent operation \cite{PRA93.032336}. According to Example 3, $M(\mathcal{E})=|\alpha^{2}-\beta^{2}|=\sqrt{1-C_{l_{1}}^{2}(|\psi\rangle)}$, where $C_{l_{1}}$ is a coherence measure based on the $l_{1}$ norm \cite{PRL113.140401}, and equals the coherence concurrence for the qubit case \cite{PRA92.022124,JPA50.285301}.

\emph{Example 5}~~Consider the quantum ensemble $\mathcal{E}=\{(1/2,|\psi\rangle),(1/2,|\phi^{+}\rangle)\}$, where $|\psi\rangle$ is a two-qubit pure state with the Schmidt decomposition $|\psi\rangle=\alpha|00\rangle+\beta|11\rangle$, $|\phi^{+}\rangle=(|00\rangle+|11\rangle)/\sqrt{2}$. According to Example 3, $M(\mathcal{E})=|\alpha^{2}-\beta^{2}|=\sqrt{1-C^{2}(|\psi\rangle)}$, where $C(|\psi\rangle)$ is the famous entanglement measure--concurrence \cite{PRL78.5022,PRL80.2245}.

\emph{Example 6}~~ We define quantity $D(\rho^{ab}):=M(\mathcal{E})$ for a classical-quantum state $\rho^{ab}=\sum_{i}p_{i}|i\rangle\langle i|\otimes\rho_{i}$, which corresponds to ensemble $\mathcal{E}=\{(p_{i},\rho_{i})|i\in I\}$. Evidently, $D(\rho^{ab})=0$ is equivalent to $D_{b}(\rho^{ab})=0$ (i.e., $\rho^{ab}$ is a classically correlated state). $D_{b}(\rho^{ab})$ denotes the quantum discord (with von Neumann measurements performed on subsystem $b$) \cite{PRL88.017901,JPA34.6899}. $D(\rho^{ab})$ possesses the following properties: (1) it has positivity; (2) it vanishes for the classical--classical state; (3) it is invariant under local unitary transformations; and (4) it is non-increasing when an ancillary system is introduced. Consequently, the quantity $M(\mathcal{E})$ may also be viewed as an easily computed measure of quantum correlation, which is different from the previous measures of quantum correlation. The previous measures are evidently hard to evaluate because a highly complicated optimization must be considered.

\section{Conclusion}
We proposed herein a class of measures using unitary similarity invariant norms of the commutators of the constituent density operators of a quantum ensemble to quantify the quantumness in a general quantum ensemble. These measures possessed desirable properties for a quantifier of the quantumness of an ensemble. We also revealed the rich applications of these measures of quantumness in different branches of quantum information theory by considering some concrete examples. For example, we established the relationship between quantumness of ensembles and quantum coherence, quantum entanglement, and quantum correlation in a simple quantum system. Our findings provide a new method of understanding these quantum resources from the viewpoint of quantumness of an ensemble. Therefore, further research on the potential link between them would be interesting and meaningful for a more complicated case. The issue of quantifying the quantumness of quantum ensembles plays a fundamental role in quantum foundations and quantum information. We hope that our results will be beneficial to the study of the quantumness of ensembles.

\begin{acknowledgments}
This work was supported by the National Natural Science Foundation of China under grant nos. 11371005 and 11475054 and the Hebei Natural Science Foundation of China under grant nos. A2016205145 and A2018205125.
\end{acknowledgments}


\begin{thebibliography}{99}

\bibitem{QCI2010} M. A. Nielsen and I. L. Chuang, \textit{Quantum Computation and Quantum Information} (Cambridge University Press, Cambridge,  England, 2010).

\bibitem{QI2011} L. Di\'{o}si, \textit{A Short Course in Quantum Information Theory: An Approach From Theoretical Physics}, 2nd ed. (Springer-Verlag, Berlin Heidelberg, 2011).

\bibitem{PRL74.1259} S. Massar and S. Popescu, Optimal extraction of information from finite quantum ensembles, \href{http://dx.doi.org/10.1103/PhysRevLett.74.1259}{Phys. Rev. Lett. \textbf{74}, 1259 (1995).}

\bibitem{QCCM2002} C. A. Fuchs, Just two nonorthogonal quantum states, \textit{Quantum Communication, Computing, and Measurement} \textbf{2}, 11-16  (Springer, Boston, MA, 2002).

\bibitem{QIC2003} C. A. Fuchs and M. Sasaki, Squeezing quantum information through a classical channel: measuring the quantumness of a set of quantum states, \href{http://www.rintonpress.com/journals/qiconline.html#v3n5}{Quantum Inf. Comput. \textbf{3}, 377 (2003).}

\bibitem{0302108} C. A. Fuchs and M. Sasaki, The quantumness of a set of quantum states, \href{https://arxiv.org/abs/quant-ph/0302108}{arXiv:quant-ph/0302108.}


\bibitem{IJQI2006} M. Horodecki, P. Horodecki, R. Horodecki, and M. Piani, Quantumness of ensemble from no-broadcasting principle, \href{https://doi.org/10.1142/S0219749906001748}{Int. J. Quantum Inform. \textbf{4}, 105 (2006).}

\bibitem{PRA79.032336} O. Oreshkov and J. Calsamiglia, Distinguishability measures between ensembles of quantum states, \href{http://doi.org/10.1103/PhysRevA.79.032336}{Phys. Rev. A \textbf{79}, 032336 (2009).}


\bibitem{PLA2011} X. Zhu, S. Pang, S. Wu, and Q. Liu, The classicality and quantumness of a quantum ensemble, \href{http://dx.doi.org/10.1016/j.physleta.2011.03.038}{Phys. Lett. A \textbf{375}, 1855 (2011).}


\bibitem{SR2014} T. Ma, M. J. Zhao, Y. K. Wang, and S. M. Fei, Non-commutativity and local indistinguishability of quantum states, \href{http://doi.org/10.1038/srep06336}{Sci. Rep.  \textbf{4}, 6336 (2014).}




\bibitem{PMH2009} S. Luo, N. Li, and X. Cao, Relative entropy between quantum ensembles, \href{http://doi.org/10.1007/s10998-009-0223-7}{Periodica Mathematica Hungarica \textbf{59}, 223 (2009).}

\bibitem{QIP2010} S. Luo, N. Li, and W. Sun, How quantum is a quantum ensemble, \href{http://doi.org/10.1007/s11128-010-0162-5}{Quantum Inf. Process. \textbf{9}, 711 (2010).}

\bibitem{TMP2011} S. Luo, N. Li, and S. Fu, Quantumness of quantum ensembles, \href{http://dx.doi.org/10.1007/s11232-011-0147-2}{Theor. Math. Phys.  \textbf{169}, 1724 (2011).}

\bibitem{PRA96.022132} N. Li, S. Luo, and Y. Mao, Quantifying the quantumness of ensembles, \href{http://doi.org/10.1103/PhysRevA.96.022132}{Phys. Rev. A \textbf{96}, 022132 (2017).}

\bibitem{PRL105.190502} B. Daki\'{c}, V. Vedral, and \v{C}. Brukner, Necessary and sufficient condition for nonzero quantum discord, \href{http://doi.org/10.1103/PhysRevLett.105.190502}{Phys. Rev. Lett. \textbf{105}, 190502 (2010).}

\bibitem{PRL106.120401} S. Luo and S. Fu, Measurement-induced nonlocality, \href{http://doi.org/10.1103/PhysRevLett.106.120401}{Phys. Rev. Lett. \textbf{106}, 120401 (2011).}

\bibitem{NJP2015} M. L. Hu and H. Fan, Measurement-induced nonlocality based on trace norm, \href{http://doi.org/10.1088/1367-2630/17/3/033004}{New J. Phys. \textbf{17}, 033004 (2015).}

\bibitem{Zhan.2013} X. Zhan, \textit{Matrix Theory}, Graduate Studies in Mathematics Vol. 147 (American Mathematical Society, Providence, Rhode Island, 2013).

\bibitem{PRA93.032336} Y. Peng, Y. Jiang, and H. Fan, Maximally coherent states and coherence-preserving operations, \href{http://dx.doi.org/10.1103/PhysRevA.93.032326}{Phys. Rev. A \textbf{93}, 032336 (2016).}


\bibitem{PRL113.140401} T. Baumgratz, M. Cramer, and M. B. Plenio, Quantifying coherence, \href{http://dx.doi.org/10.1103/PhysRevLett.113.140401}{ Phys. Rev. Lett. \textbf{113}, 140401 (2014).}

\bibitem{PRA92.022124} X. Yuan, H. Zhou, Z. Cao, and X. Ma, Intrinsic randomness as a measure of quantum coherence, \href{http://dx.doi.org/10.1103/PhysRevA.92.022124}{Phys. Rev. A \textbf{92}, 022124 (2015).}

\bibitem{JPA50.285301} X. F. Qi, T. Gao, and F. L. Yan, Measuring coherence with entanglement concurrence,  \href{https://doi.org/10.1088/1751-8121/aa7638}{J. Phys. A: Math. Theor. \textbf{50}, 285301 (2017).}

\bibitem{PRL78.5022} S. Hill and W. K. Wootters, Entanglement of a pair of quantum bits,  \href{http://doi.org/10.1103/PhysRevLett.78.5022}{Phys. Rev. Lett. \textbf{78}, 5022 (1997).}

\bibitem{PRL80.2245} W. K. Wootters, Entanglement of formation of an arbitrary state of two qubits, \href{http://doi.org/10.1103/PhysRevLett.80.2245}{Phys. Rev. Lett. \textbf{80}, 2245 (1998).}

\bibitem{PRL88.017901} H. Ollivier and W. H. Zurek, Quantum discord: a measure of the quantumness of correlations, \href{http://dx.doi.org/10.1103/PhysRevLett.88.017901}{ Phys. Rev. Lett. \textbf{88}, 017901 (2001).}

\bibitem{JPA34.6899} L. Henderson and V. Vedral, Classical, quantum and total correlations, \href{https://doi.org/10.1088/0305-4470/34/35/315}{J. Phys. A: Math. Gen. \textbf{34}, 6899 (2001).}

\end{thebibliography}
\end{document}